\documentclass{pasj01}
\draft 
\Received{$\langle$reception date$\rangle$}
\Accepted{$\langle$acception date$\rangle$}
\Published{$\langle$publication date$\rangle$}

\usepackage{graphics,epsfig,graphicx}
\usepackage{natbib}
\usepackage{lscape}

\def\spose#1{\hbox to 0pt{#1\hss}}
\def\lta{\mathrel{\spose{\lower 3pt\hbox{$\mathchar"218$}}
     \raise 2.0pt\hbox{$\mathchar"13C$}}}
\def\gta{\mathrel{\spose{\lower 3pt\hbox{$\mathchar"218$}}
     \raise 2.0pt\hbox{$\mathchar"13E$}}}

\begin{document}

\title{Can Planets Exist in the Habitable Zone of 55~Cancri?}
\author{\textsc{Suman Satyal}\altaffilmark{1}%
}
\altaffiltext{1}{Department of Physics, University of Texas at Arlington, Box 19059,
Arlington, TX 76019, USA}
\email{suman.satyal@uta.edu}

\author{\textsc{Manfred Cuntz}\altaffilmark{1}}
\email{cuntz@uta.edu}

\KeyWords{astrobiology --- celestial mechanics --- methods: numerical ---
planetary systems --- stars: individual (55~Cnc)}

\maketitle

\begin{abstract}
The aim of our study is to explore the possible existence of Earth-mass planets in the
habitable zone of 55~Cancri, an effort pursued based on detailed orbital stability simulations.
This star is known to possess (at least) five planets with masses ranging between super-Earth
and Jupiter-type.  Additionally, according to observational constraints, there is a space without
planets between $\sim$0.8~au and $\sim$5.7~au, noting that the inner part of this gap largely
coincides with 55~Cnc's habitable zone --- a sincere motivation for the search of potentially
habitable planets.  It has previously been argued that terrestrial habitable planets are able
to exist in the 55~Cnc system, including a planet at $\sim$1.5~au.  We explore this possibility
through employing sets of orbital integrations and assuming an integration time of 50~Myr.
We found that the possibility of Earth-mass planets in the system's habitable zone strongly
depends on the adopted system parameters, notably the eccentricity of 55~Cnc-f, which is
controversial as both a high value ($e \sim 0.32$) and a low value ($e \sim 0.08$) have
previously been deduced.  In case that the low value is adopted (together with other updates
for the system parameters), the more plausible and most recent value, Earth-mass planets would
be able to exist in the gap between 1.0~au and 2.0~au, thus implying the possibility of habitable
system planets.  Thus, 55~Cnc should be considered a favorable target for future
habitable planet search missions.
\end{abstract}


\section{Introduction}

A prevalent topic at the crossroad between astrophysics and astrobiology is given by the
study of stellar habitable zones (HZ).  Following the landmark paper by \cite{kas93}, which focused on
the extent of HZs for main-sequence stars based on simplified atmospheric climate models two and a
half decades ago, there has been an explosion of literature on that topic.  Immense progress has been
made especially in the study of planetary atmospheres, see, e.g., \cite{kop13,kas15,ram18}, among many
other studies.  Recently, a catalog of HZ exoplanet candidates based on results of the {\it Kepler} mission
has been published as well by \cite{kan16}.  Clearly, the progress made in investigating HZ also goes
hand-in-hand with ongoing discoveries of extrasolar planets around different types of stars\footnote{For
updated information, please visit {\tt http://exoplanet.eu}}, including G and K-type stars.
However, it is well-acknowledged that the location of a planet inside the HZ is by itself insufficient
for ensuring habitability, as sets of other conditions need to be fulfilled, including (but not limited to)
the mass and size of that planets, geodynamic aspects, and environmental forcings mostly associated with
the parent star; see, e.g., \cite{lam09}, and subsequent work.

Regarding astrobiology, orange dwarfs, i.e., late-type G and early-type K stars have previously
received a heightened amount of attention \citep[e.g.,][]{hel14,cun16}.  These types of stars are
considered particularly suitable for hosting planets with exolife (potentially
even advanced forms of life) due to numerous supportive features including the frequency of those stars,
the relatively large size of their HZs (if compared to M-type stars)\footnote{This aspect has recently
been challenged as observations and theoretical work show that the occurrence rate of low-mass planets
increases with smaller orbital distances from the star, thus favoring the population of HZs of very
low-mass stars; see \cite{mul18} and \cite{ogi18}.  However, this finding does not reduce the
privileged situation of orange dwarfs relative to other types of stars as they are advantageous
to hotter stars, but not disadvantageous to cooler stars such as M dwarfs (although those
may host an increased number of low-mass planets in their HZs), noting that the latter are often
considered subideal for hosting habitable planets owing to their energetic radiative environments.}
their relatively long life time on the main-sequence (i.e., 15~Gyr to 30~Gyr, compared to $\sim$10~Gyr for
stars akin to the Sun), and the modest amounts of magnetic-dynamo-generated X-ray-UV emissions
(except young stars) resulting in not-so-much destructive planetary atmospheric environments; see,
e.g., \cite{cun16} and references therein.

In this work, we focus on 55~Cancri (55~Cnc, $\rho^1$~Cnc), a G8~V star \citep{gon98}, with
an effective temperature and luminosity lower than those of the Sun \citep[e.g.,][]{fis05,lig16}.  Moreover,
55~Cnc is considerably older than the Sun with an estimated age between 7.4~Gyr and 8.7~Gyr \citep{mam08}.
It has a mass of 0.96~$M_\odot$ \citep{lig16}.
Additionally, 55~Cnc is known to host five planets, discovered between 2008 and 2011 by \cite{fis08},
\cite{daw08}, and \cite{win11} based on the radial velocity method.  All of these planets have masses
significantly larger than Earth.  However, based on previous theoretical studies, see, e.g.,
\cite{blo03}, \cite{riv07}, \cite{ray08}, and \cite{smi09}, the general possibility of Earth-mass
planets in that system is implied.

The focus of this study is to explore whether 55~Cnc would be able to host Earth-mass planets,
especially in its HZ.  This topic is of interest to both astrophysics and astrobiology, and
is expected to be relevant to future planetary search missions.  Thus, we will employ detailed
orbital stability simulations for Earth-mass test planets assuming adequate initial conditions (ICs).
In this respect, we consider previous data for this system, as well as updated findings by
\cite{bou18}.  Regarding those studies there is a pivotal difference in the eccentricity
of 55~Cnc-f, which is of critical importance for the outcome of our simulations.
Our paper is structured as follows.  In Section~2, we convey our theoretical approach, including
information on the system set up as well as the theoretical simulations.  Our results and discussion
are given in Section 3.  Comments on previous studies are given in Section 4, whereas Section 5
states our summary and conclusions.


\section{Theoretical Approach}

\subsection{System Set-Up}

The star 55~Cnc is host to (at least) five planets, discovered between 2008 and 2011, with masses ranging from super-Earth to Jupiter-type; see Table~1 for details and references\footnote{In this study, we will assume minimum masses for the five 55~Cnc system planets.  If higher masses were assumed, the domain of orbital stability for hypothetical Earth-mass planets would be further reduced.}.
The location of the planets in the 55~Cnc system, as identified, exhibits a relatively large gap between 55~Cnc-f and 55~Cnc-d.  Therefore, to numerically test if any additional planet could remain in a stable orbit in this region, we inject Earth-mass planets with different ICs and perform dynamical analyses.  The integrations are carried out to determine the nearest stable semi-major axis exterior to 55~Cnc-f, which is at $\sim$0.77~au.  Injected planets are set at initially circular and coplanar orbits
at distances of $a_{\rm pl}=a_0$.
Next, the planetary initial inclination ($i_{\rm pl}=i_0$) and initial eccentricity ($e_{\rm pl}=e_0$) are gradually increased to explore different dynamical settings.  However, for the inclination no values above 20$^\circ$
are considered as no significant dynamical changes were found to occur in the region-of-interest.  Compact systems like 55~Cnc are typically identified as having nearly coplanar orbits.  For example, the planets of the Kepler-62 and TRAPPIST-1 systems have inclinations varying by less than 1$^\circ$ \citep{bor13,gil16}.  However, some of these data may have been biased by the applied observation method.

The initial orbital parameters, i.e., semi-major axis ($a_0$), eccentricity ($e_0$), inclination ($i_0$),
argument of periapsis ($\omega_0$), ascending node ($\Omega_0$), and mean anomaly ($MA_0$) of the five planets are obtained from
the original discovery papers and follow-up papers, if applicable; see \cite{fis08}, 
\cite{daw08}, \cite{win11}, \cite{end12}, and \cite{lig16}.  These parameters are given in Table~1.  We also considered a different set of orbital parameters recently reported by \cite{bou18}; they differ significantly from the previous works (see Table~2).  We will present and discuss the outcome of simulations from both sets of data.  However, we will not consider the observational uncertainties in these parameters as part of our simulation; the latter are based on the planets' best-fit values.  The orbital parameters, i.e., $\omega$, $\Omega$, and $MA$, not reported in the respective articles are set to zero.

Technically, 55~Cancri constitutes a wide binary system.  55~Cancri~A, the focus of our study, is accompanied by a red dwarf, 55~Cancri~B, with a mass of 
0.13~$M_\odot$.  Due to the large separation of the secondary companion, given as $\sim$1065~au \citep{duq91,egg04}, there is no need to consider its gravitational effects on the planets for our integrations\footnote{Methods for the calculation of HZs for 55~Cancri-type binaries have been given by, e.g., \cite{egg13} and \cite{cun14,cun15}.}.

\subsection{Numerical Simulations}

The numerical calculations are performed using the orbital integration package \texttt{mercury} \citep{cha97,cha99} where the multi-body system is set in an astro-centric coordinate system.  The program calculates the orbital evolution of the planets as they orbit in the gravitational field given by the central star and the five system planets.  Among different N-body algorithms within the program, the Hybrid Symplectic/Bulirsch-Stoer Integrator has been chosen because of its fastness and its ability of computing close encounters.  In the \texttt{mercury} setup files, the time step is fixed at $\epsilon = 10^{-3}$ year/step in order to minimize the error accumulation. The 7-body system (5~planets as observed, 1~Earth-mass test planet, and the star) is simulated for up to 50 Myr.  The data are sampled every 1000 years for short-term simulations and every 10,000 years for long-term simulations.  Additionally, we monitor the
changes in the system's total energy and angular momentum; both of them consistently remain close to zero.  Other forces as, e.g., tides are not considered as part of the simulations.

For the initial orbital parameters of the test planets, the semi-major axis is assumed to take values between 0.0~au and 6.0~au with a step size of 0.01~au. First, we consider coplanar and circular orbits.  Thereafter, the test planets are integrated for non-coplanar orbits by varying the orbital inclination from 0$^\circ$ to 20$^\circ$ with a step size of 0.5$^\circ$, and the orbital eccentricity from 0.0 to 0.4 with a step size of 0.01.  Hence, a total of 24,000 ICs are simulated in the $a_{\rm pl}$ - $i_{\rm pl}$ phase space, and 20,000 ICs are simulated in the $a_{\rm pl}$ - $e_{\rm pl}$ phase space. The mass of the test planets is always set to 1~$M_\oplus$; furthermore, the other orbital parameters (i.e., argument of periapsis, ascending node, and mean anomaly) are set to zero in all models.  The relevant parameters, including those for the test planets
are given in Table \ref{tab:planets_param} for earlier work and Table \ref{tab:planets_param_new} based on work by \cite{bou18}; see discussion below.

The orbits of the Earth-mass test planets were considered stable if they survived the total simulation time and their maximum eccentricities remained close to their initial values, without approaching $\sim$1. In the cases where the test planets collided with one of the system planets or the star, or if they were ejected from the system, the orbits were considered unstable. For each simulation, this information can be extracted via \texttt{mercury} output files for further analyses.


\section{Results and Discussion}

\subsection{Stability Analysis Based on Earlier Work}

In the following, we report on numerical results based on data given in Table~1, which encompasses results published between 2008 and 2016.
Generally, an important aspect pertains to the comparison of the domain of planetary orbital stability to the stellar HZ.
The HZ of 55~Cnc is calculated using the formalism given by \citet{kop13,kop14}.  They have specified a general habitable zone (GHZ) as well as an optimistic habitable zone (OHZ); the latter is given by the recent Venus / early Mars limits, previously coined by \cite{kas93}. Based on the luminosity and stellar effective temperature of 55 Cnc, given as $L=0.589~L_\odot$ and $T_{\rm eff}=5165$~K, respectively, see \cite{lig16}, the GHZ extends from $\sim$0.78~au to $\sim$1.37~au, and the OHZ extends from $\sim$0.59~au to $\sim$1.43~au. Hence, our analysis is concentrated at distances akin to the GHZ and OHZ, as well as at distances close to $\sim$1.51~au previously advocated by \cite{cun12}. Regions akin to those distances will be selected to inject hypothetical Earth-mass planets in order to explore their orbital stability.

\begin{figure*}
\begin{center}
     \includegraphics[scale=.270,angle=0]{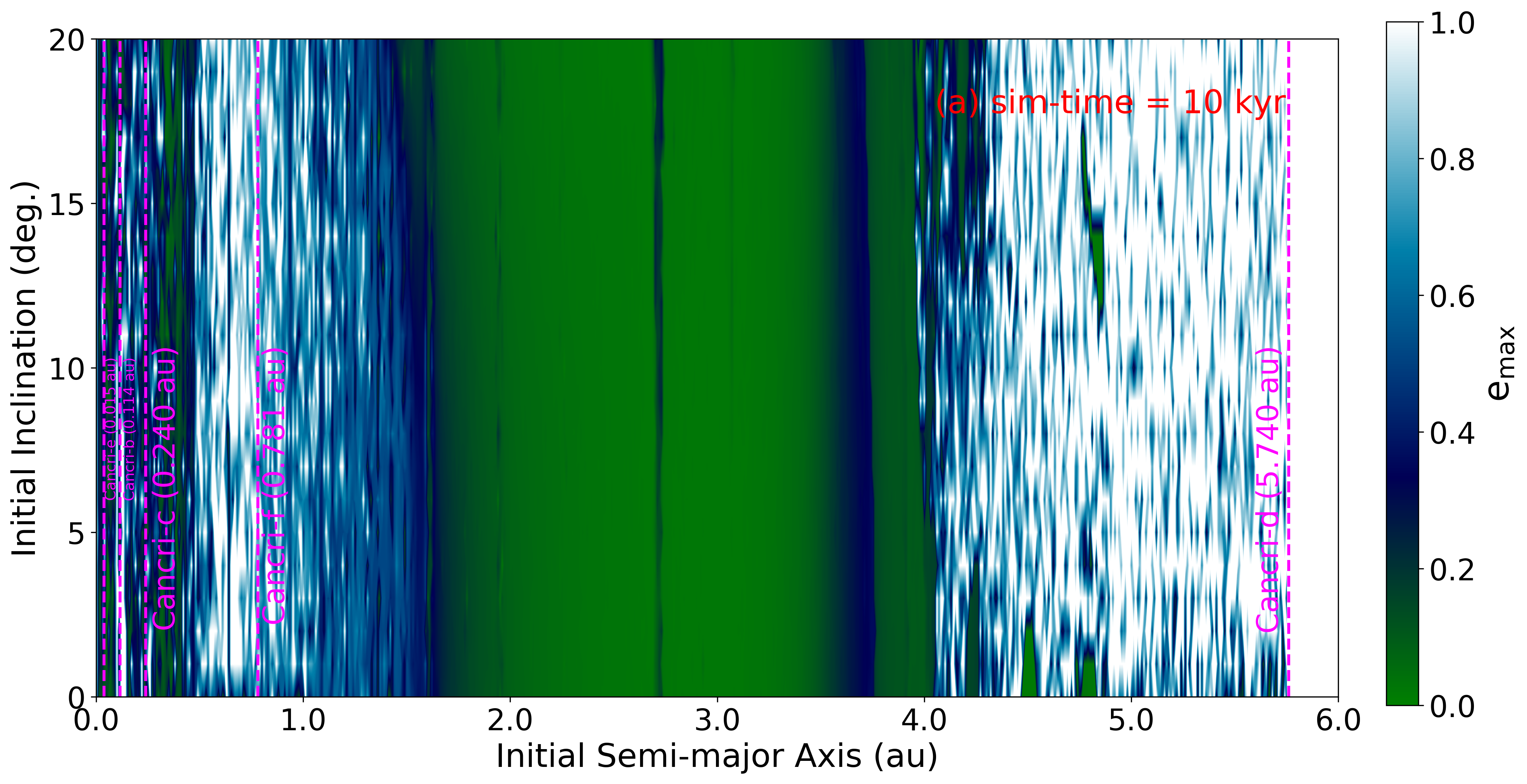} 
	\includegraphics[scale=.270,angle=0]{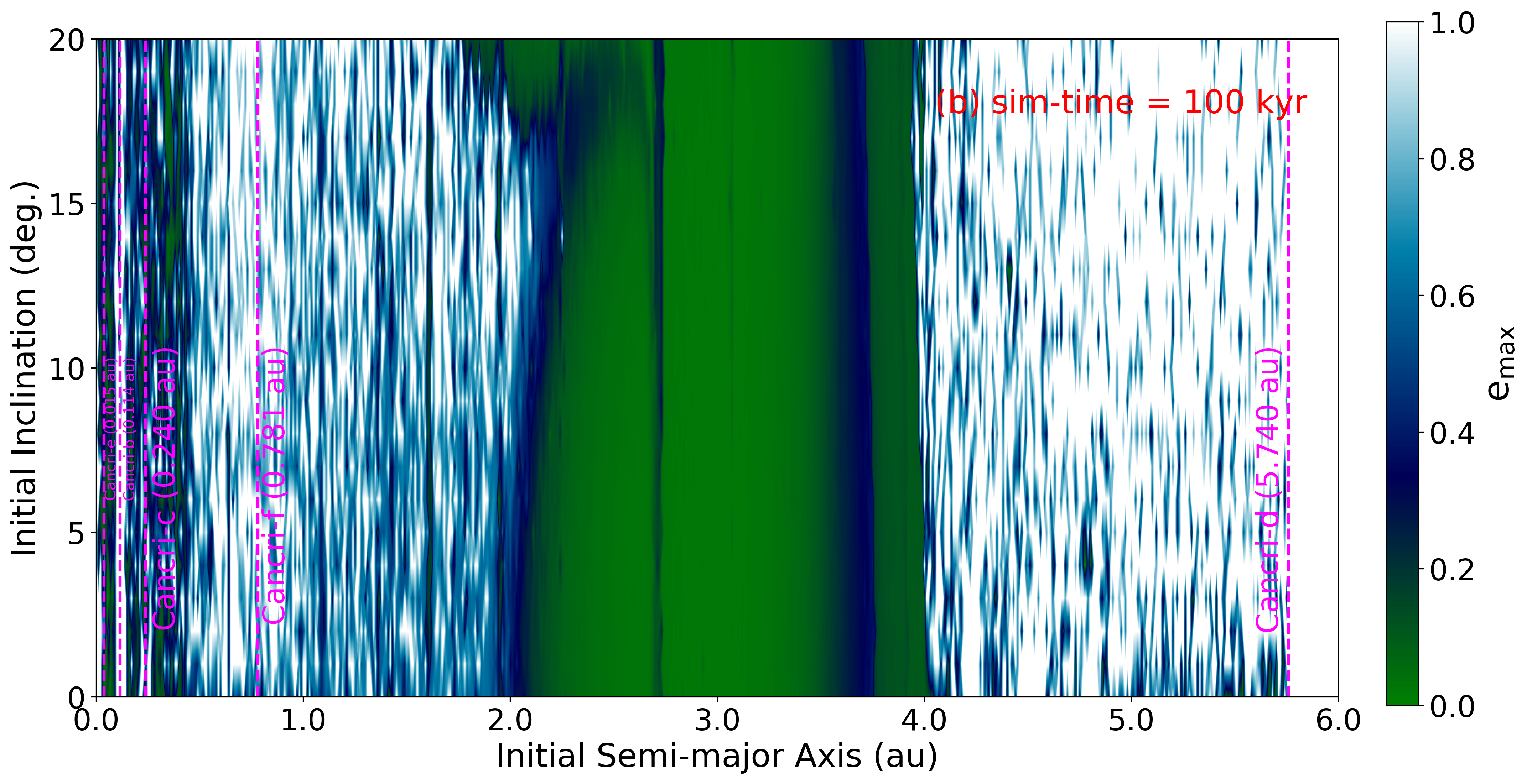} \\
	\includegraphics[scale=.270,angle=0]{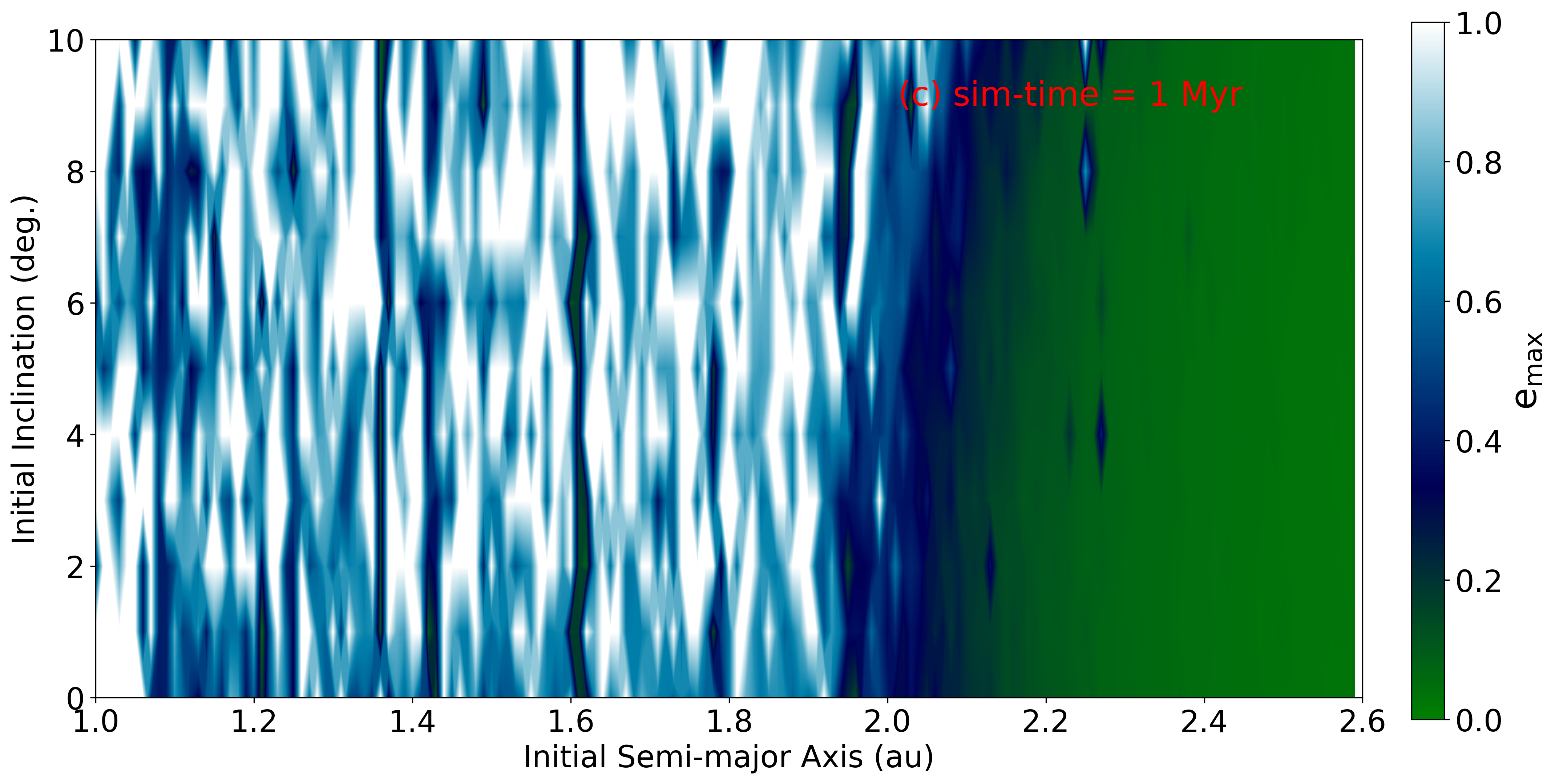}
	\includegraphics[scale=.270,angle=0]{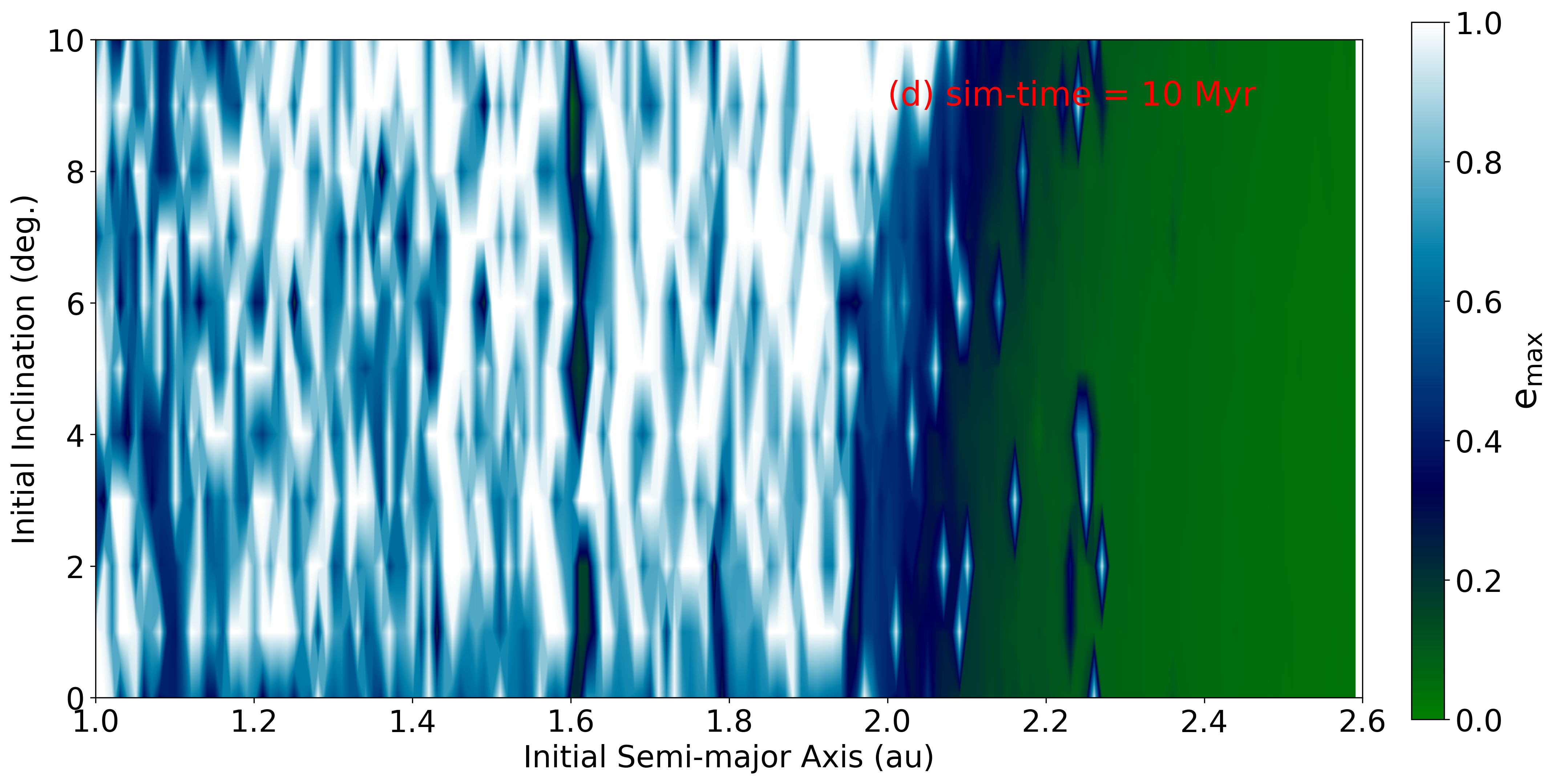}
\caption{Maximum eccentricity ($e_{\rm max}$) of the Earth-mass test planets that are distributed between 0.0~au and 6.0~au
(i.e., mostly the spatial region extending from the innermost to the outermost system planet).  The vertical magenta lines in (a) and (b) represent
the location of the five system planets.  The plots (a, b, c, d) depict the $e_{\rm max}$ values for simulation times
of 10~kyr, 100~kyr, 1~Myr, and 10~Myr, respectively.  In all panels the $e_{\rm max}$ values represented by the green color
mean that the Earth-mass planets stay close to zero for the total simulation time.  When a value transitions from dark-blue
to light-blue to white, it indicates that the $e_{\rm max}$ value approaches 1.0, and the orbit of the Earth-mass planet becomes
unstable.}
\label{fig:1}
\end{center}
\end{figure*}

A preliminary stability map for the injected planet based on $e_{\rm max}$ while using a relatively short simulation time of 10 kyr is given in Fig. \ref{fig:1}a.  The magenta vertical lines represent the location of the five system planets.  The color bar represents the $e_{\rm max}$ values; here green means 0.0 and white means 1.0.  Other intermediate colors change from dark green to dark blue to light blue as the eccentricity values change from 0.2 to 0.8.  The phase-space map of $i_{\rm pl}$ versus $a_{\rm pl}$ exhibits a stability region with the $e_{\rm max}$ values remaining close to zero; it extends from 1.6~au to 4.0~au. However, this stability region considerably shifts to the right from 1.6~au to 2.0~au when the simulation time is increased to 100 kyr (Fig. \ref{fig:1}b).  The unstable regions between 1.0~au and 2.0~au become more prominent for 1 Myr and 10 Myr simulation times (Fig. \ref{fig:1}c, d).  A smaller phase space (1.0~au to 2.6~au) is considered for longer simulation times to focus on the planetary behaviors in the stellar HZ and to make our simulations computationally less expensive.  A few resonance structures can be seen at 1.6~au and 1.8~au where the eccentricity values deviate the least from the initial values $e_0$; however, we expect them to disappear for simulation times larger than 10 Myr.  In the region of phase space between 1.2~au and 1.6~au, the eccentricity rises to $\sim$1 for almost all ICs.  Hence, this makes it impossible for any additional planet to remain in a dynamically stable orbit in the HZ; i.e., neither the GHZ nor the OHZ.

\begin{figure*}
\begin{center}
   \includegraphics[scale=.5,angle=0]{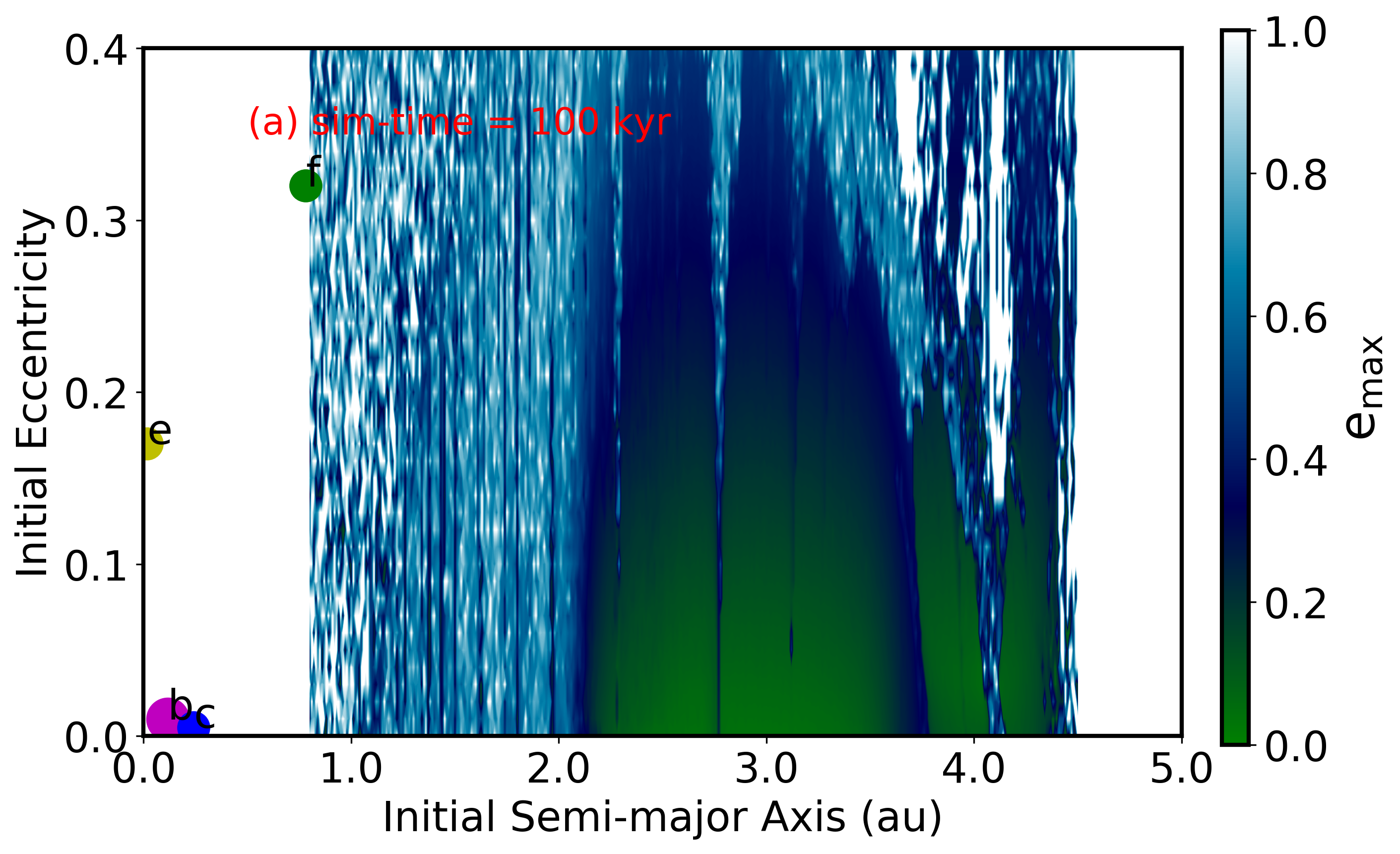}
   \includegraphics[scale=.5,angle=0]{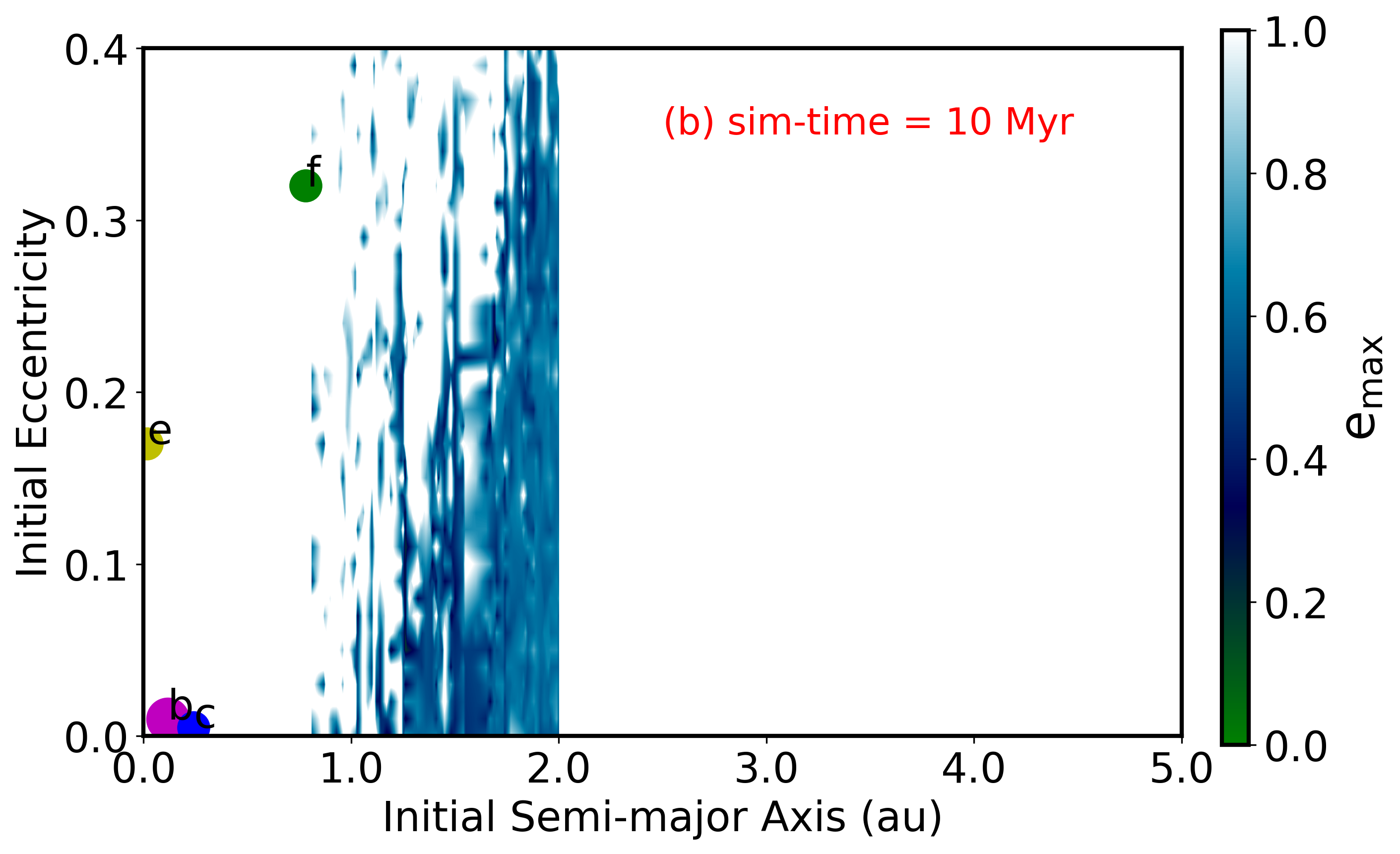}
\caption{Maximum eccentricity ($e_{\rm max}$) map for Earth-mass test planets injected between
55~Cnc-f and 55~Cnc-d in the $e_{\rm pl}$ and $a_{\rm pl}$ phase space.  The orbits are simulated
for (a) 100~kyr and (b) 10~Myr.  The index in the color bar indicates the $e_{\rm max}$ values
attained for the orbits after their dynamic evolution while starting from zero after the total
time of simulation.  The $e_{\rm pl}$ variation is relatively large between 1~au and 2~au, roughly
coinciding with the stellar HZ.  In the region between 2~au and 4~au, the eccentricity variations remain
relatively low, thus indicating stable orbits.  The green color represents
the initial value $e_{\rm pl}=e_0$
and the other colors represent the $e_{\rm max}$ value the Earth-mass planets attained for the
respective choices of ICs in $e_{\rm pl}$ and $a_{\rm pl}$.  The colored circles
labeled e, b, c, and f depict the orbital locations and sizes of four of the five system planets
in the $a_{\rm pl}$ - $e_{\rm pl}$ phase space.  The observed resonances at 2.2~au, 2.8~au, 3.1~au, 3.8~au,
and 4.12~au are due to 55~Cnc-f and 55~Cnc-d.  Similar resonances have been reported by \cite{ray08}.
However, due to the increased resolution, they seem more prominent in this work.}
\label{fig:2}
\end{center}
\end{figure*}

Figure \ref{fig:2} shows the $e_{\rm max}$ map for the injected planet with 1~$M_\oplus$, which is generated from the maximum eccentricities attained by the orbits during simulation times of 100 kyr and 10 Myr.  Four of the interior planets (Cnc-b, Cnc-c, Cnc-d, and Cnc-f) are shown by colored circles at their respective locations.  The maps indicate an unstable region between 1.0~au and 2.0~au followed by a stable region between 2.0~au and 4.0~au.  The map shows significant changes in the $e_{\rm max}$ values in the regions below 2.0~au, which partially coincides with the stellar HZ. The stability regions decrease as the simulation time is increased from 100 kyr to 10 Myr as also observed in Fig. \ref{fig:1}.  Also, similar to the Fig. \ref{fig:1}d, only a small portion of the phase space, located between 0.8~au and 2~au, is simulated for 10 Myr.  This region is expected to shrink further for simulation times in excess of 10 Myr.  However, the regions of stability around 3~au are expected to remain, and may thus be able to host additional planets.

\begin{figure*}
\begin{center}
   \includegraphics[scale=.5,angle=0]{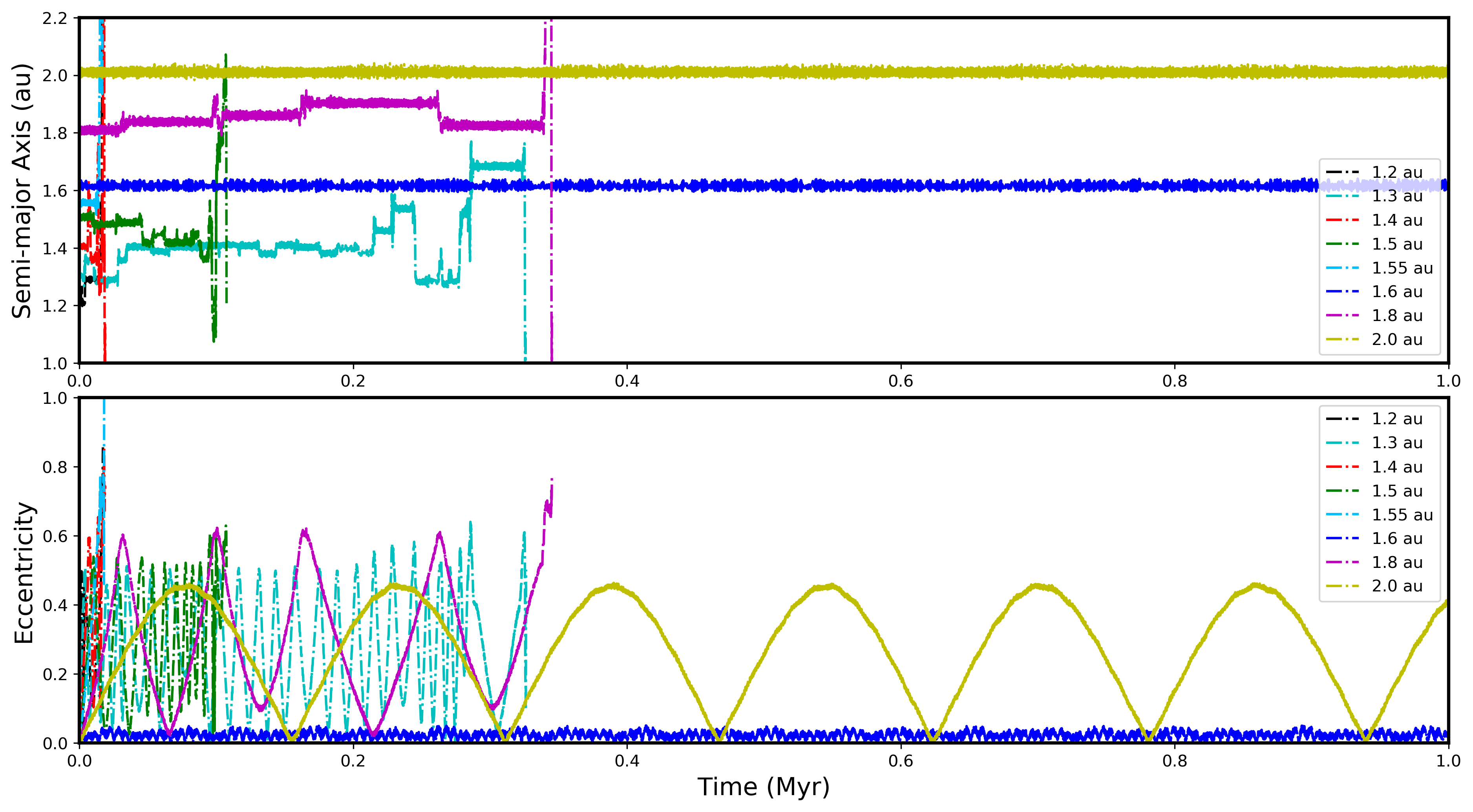}
\caption{Orbital simulations of Earth-mass test planets that are distributed between 1.0~au and 2.0 au.
The simulation time extends up to the onset of orbital instability for most of the orbits. The time series plots for semi-major axes (top panel) and the eccentricities (bottom panel) indicate the time evolution of the test planets with starting $a_{\rm pl}$ values between 1.0~au and 2.0 au. In both panels, $a_{\rm pl}$ and $e_{\rm pl}$ are found to depart from their
respective initial values $a_0$ and $e_0$, steadily increasing toward their respective instability limits.}
\label{fig:3}
\end{center}
\end{figure*}

Figure \ref{fig:3} (top panel), conveys the evolution of the semi-major axes $a_{\rm pl}$ for Earth-mass test planets originally placed between 1.0~au and 2.0~au.  All test planets in this phase space, including those originally set in the stellar HZ, become unstable by either colliding with the one of the system planets or by being ejected from the system.  The amplitude of the semi-major axis variation is rather small for some cases, like $a_{\rm pl}$ = 1.6 au and $a_{\rm pl}$ = 2 au; nevertheless, the planetary orbits become chaotic after 10 Myr when $e_{\rm pl}$ exceeds 0.5.  This behavior eventually leads to instability.  The $a_{\rm pl}$ time series is complemented by the bottom panel depicting the evolution of $e_{\rm pl}$ for the same planetary orbits.  Note that the amplitude of $e_{\rm pl}$ oscillations varies significantly between 0.0 and 1.0 for all cases, which indicates that the orbits will eventually become unstable.  The test planets set at distances below 1.5~au also became orbitally unstable during a relatively short simulation time due to collisions with the system planets.  Survival times and $e_{\rm max}$ values for all test planets set between 1.5~au and 1.6~au are given in Table 3.

\begin{figure*}
\begin{center}
  \includegraphics[scale=.35,angle=0]{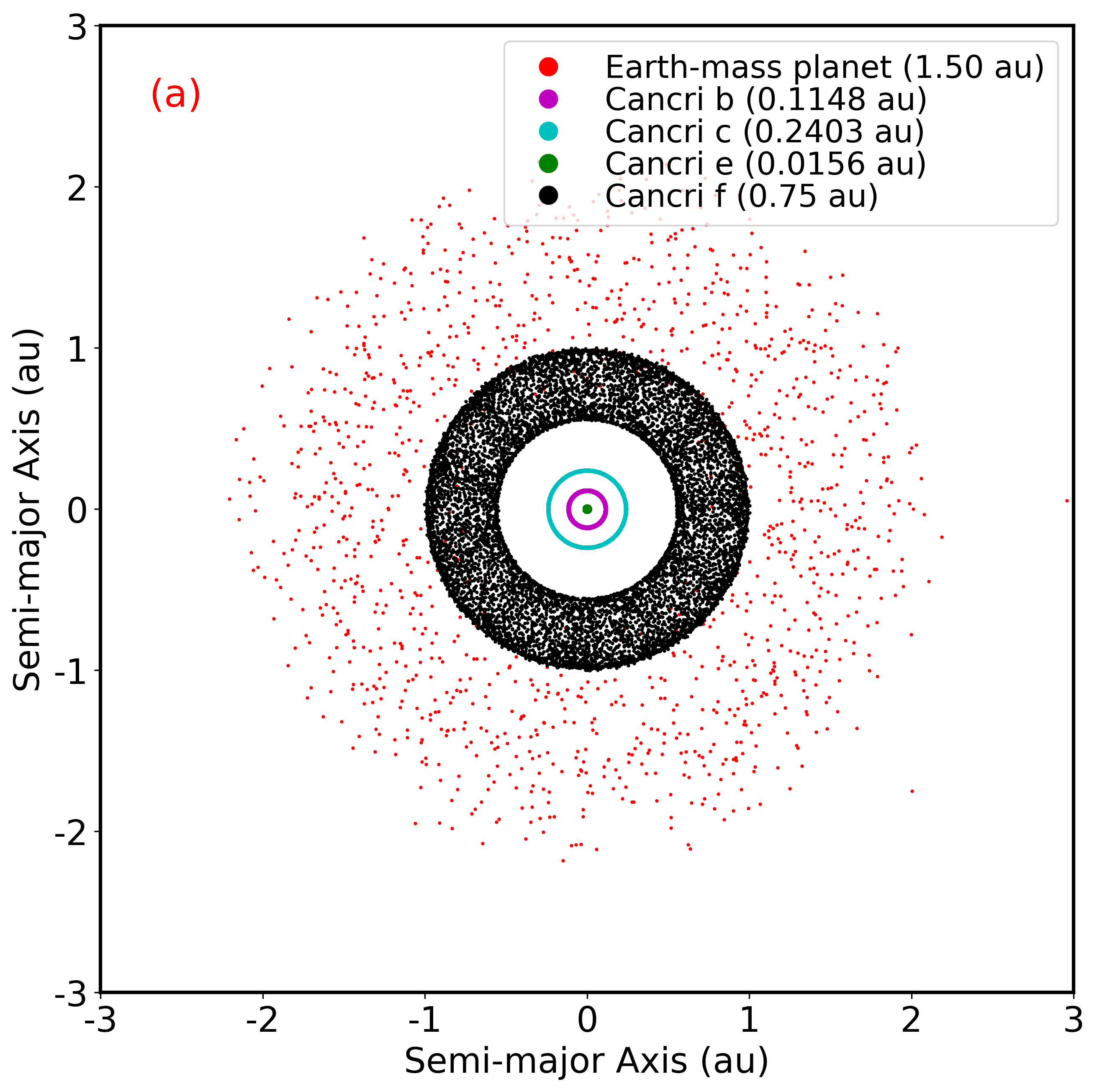}
  \includegraphics[scale=.35,angle=0]{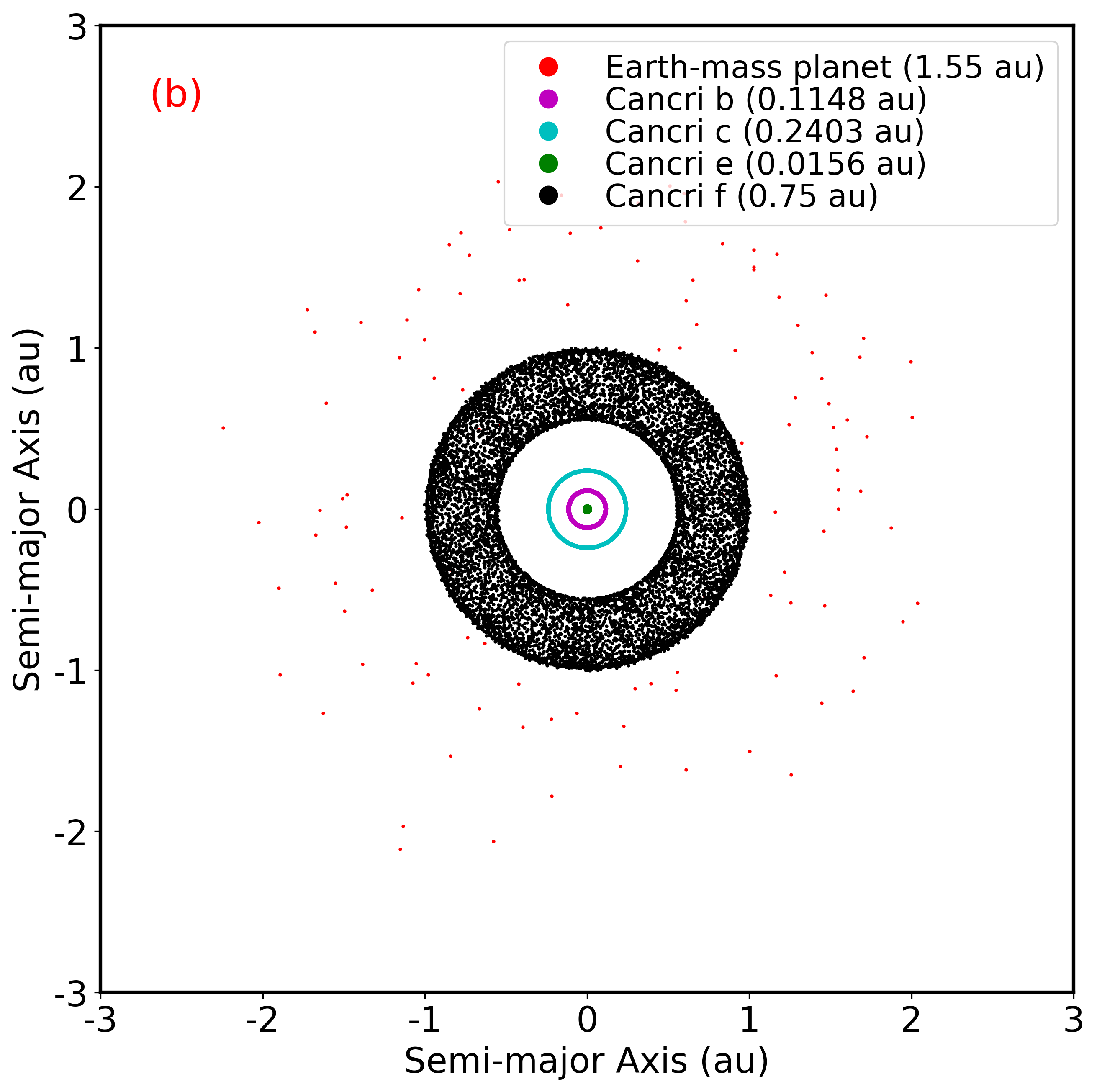} \\
  \includegraphics[scale=.35,angle=0]{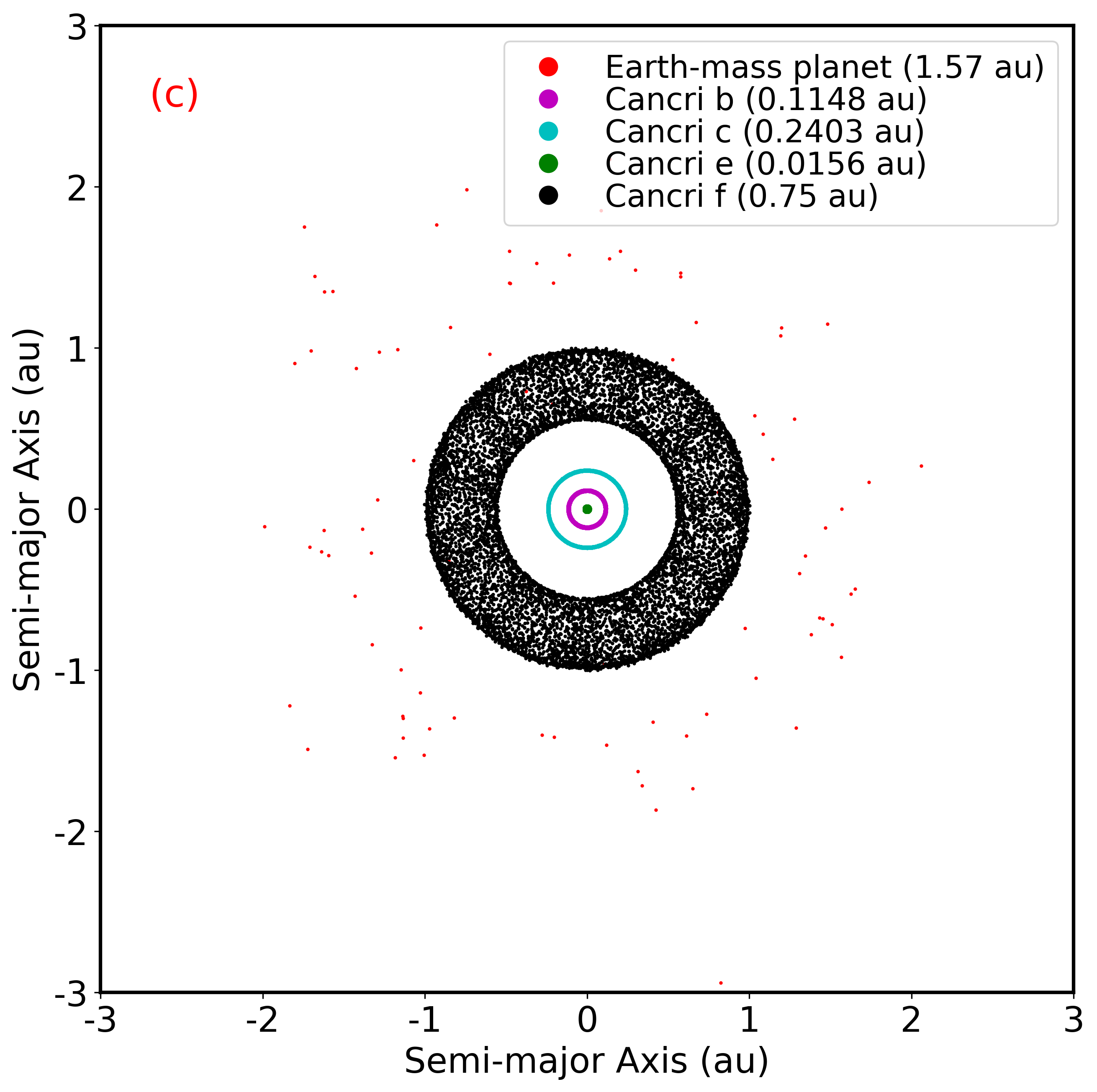}
  \includegraphics[scale=.35,angle=0]{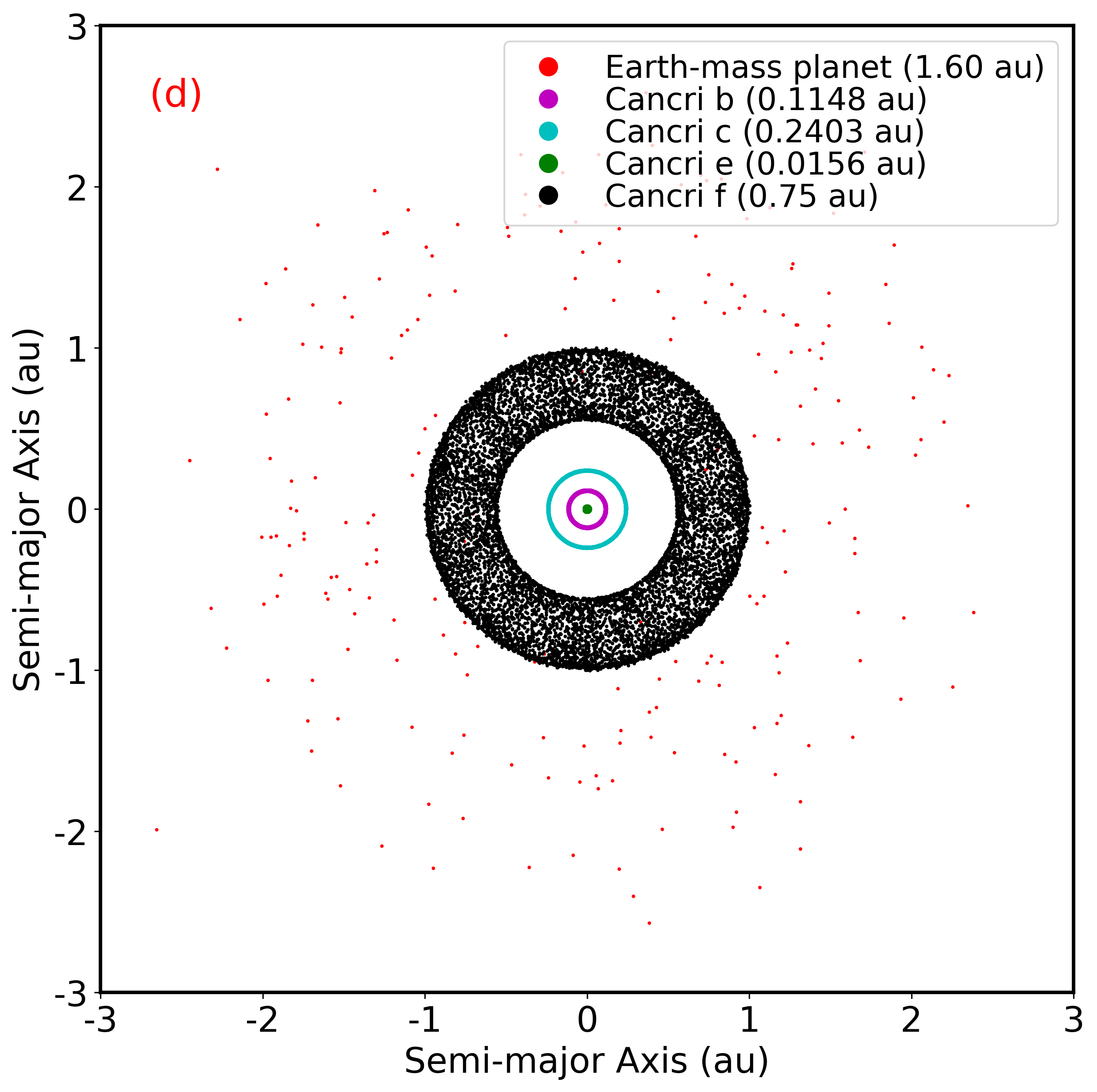}
\caption{Two-dimensional plots of the orbits of Earth-mass test planets (red); for
completeness, the orbits of the 55~Cancri system planets are depicted as well: 55 Cnc-b (magenta), 55 Cnc-c (cyan), 55 Cnc-d (blue), 55 Cnc-e (green), and 55 Cnc-f (black).
Results are given for Earth-mass planets originally set at (a) 1.50~au, (b) 1.55~au, (c) 1.57~au, and (d) 1.60~au. Regarding the bottom panels, 55~Cnc-d, the
outermost planet, has been omitted.  Here the simulation time extends up to the onset
of orbital instability ($\sim$ 2.5~Myr) for the added Earth-mass planets.  For the
Earth-mass planets originally located at 1.57~au and 1.60~au, respectively, their
behaviors also demonstrate the wide variation of their semi-major axes ranging between
1~au to 3~au.}
\label{fig:4}
\end{center}
\end{figure*}

Figure \ref{fig:4} depicts the orbits of the five 55~Cancri system planets, based on data from earlier work, as well as the added Earth-mass planets originally set at 1.50 au, 1.55 au, 1.57 au, and 1.60 au, respectively; see Table~\ref{tab:planets_param} for information on the data of the system planets as used.
The variations exhibited by the semi-major axis of the test planet during the simulations are relatively large.  They are observed to scatter by as much as 2~au, and are displayed in red in all four panels.  The system planets (depicted in magenta, cyan, blue, green, and black for 55 Cnc-b, 55 Cnc-c, 55 Cnc-d, 55 Cnc-e, and 55 Cnc-f, respectively) are found to exhibit periodic orbits with small variations in their semi-major axis values. The bottom two panels, i.e., for (a) $a_0$ = 1.57 au and (b) $a_0$ = 1.60 au, omit the data for 55~Cnc-d to better display the smaller inner orbits. The Earth-mass planet reaches instability within 2.5 Myr for all four $a_0$ values.  Thus, based on our results as demonstrated via phase-space maps and time series analyses (see Figs. \ref{fig:1} to \ref{fig:4}), we conclude that --- for the assumed data set --- no Earth-mass planets are able to exist at distances below $\sim$2.2~au, a domain that also encompasses the stellar HZ. Simulation times beyond 10~Myr are expected to somewhat further reduce the zone of planetary orbital stability.

\subsection{Stability Analysis Based on the Data by \cite{bou18}}

\begin{figure*}
\begin{center}
  \includegraphics[scale=.6,angle=0]{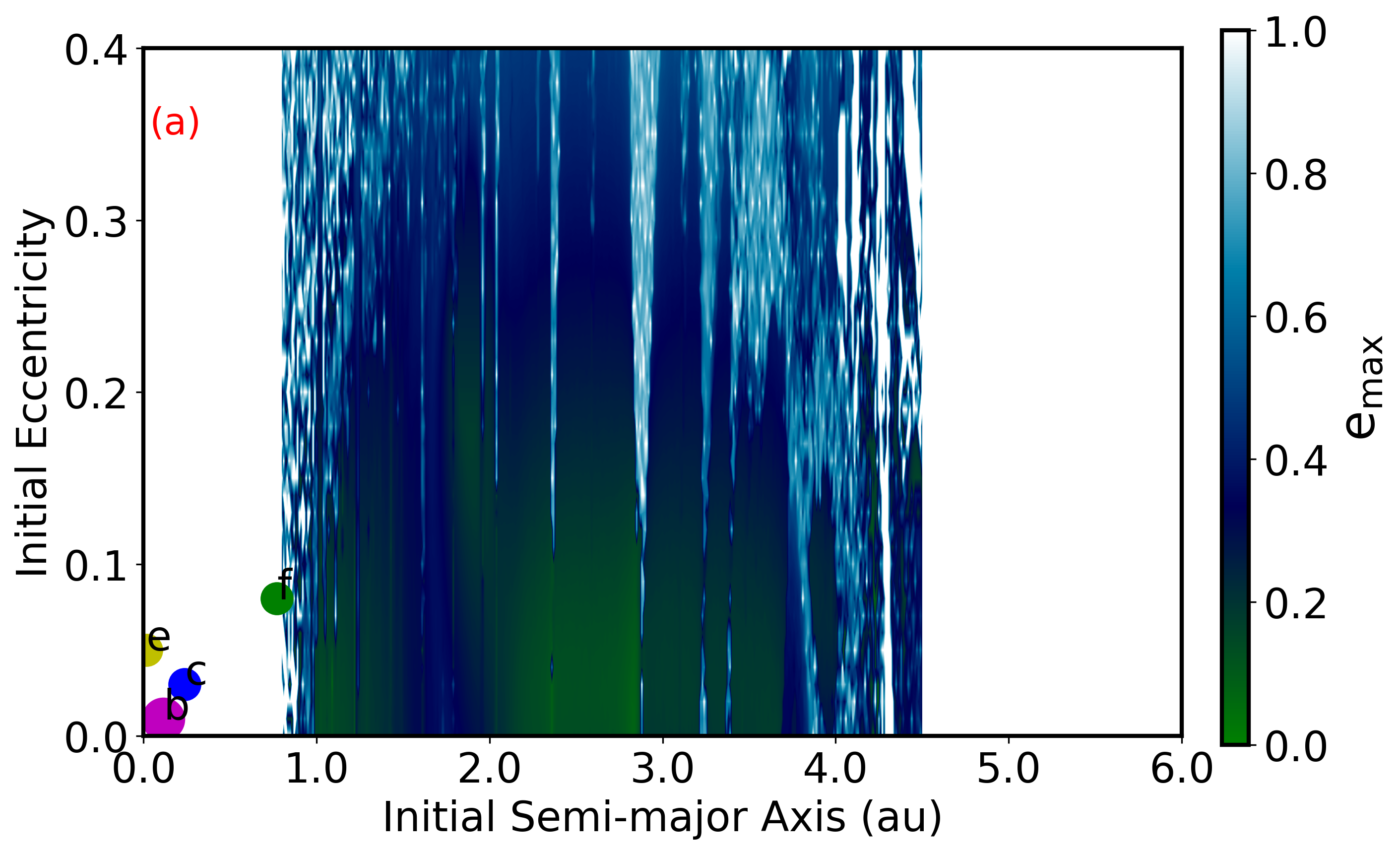}
  \includegraphics[scale=.6,angle=0]{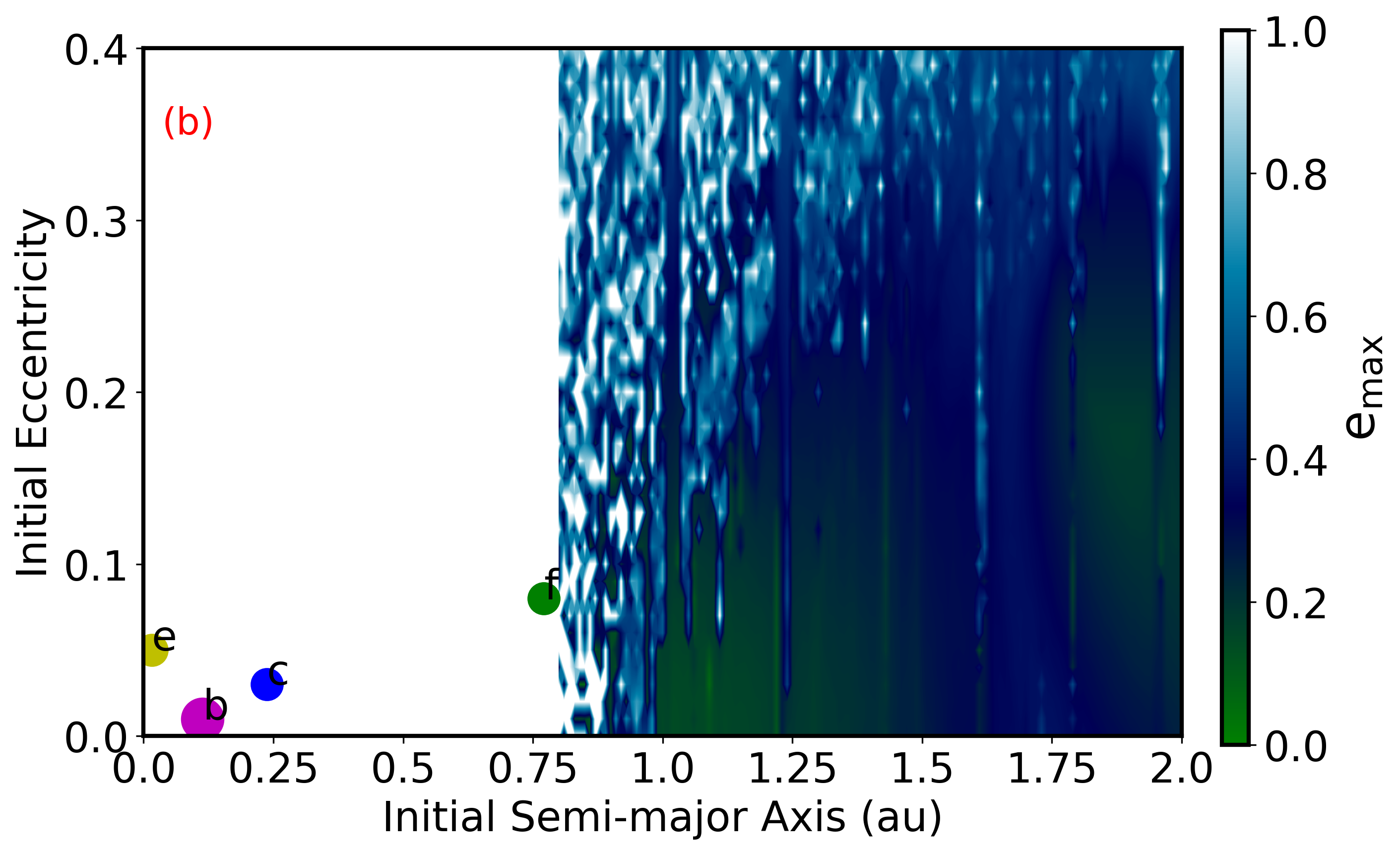} \\
\caption{Maximum eccentricity ($e_{\rm max}$) map for Earth-mass test planets injected between
55~Cnc-f and 55~Cnc-d in the $e_{\rm pl}$ and $a_{\rm pl}$ phase space. The orbits are simulated
for 100~kyr. Panel~(b) depicts the zoomed-in stability region identified in panel~(a).
The index in the color bar indicates the $e_{\rm max}$ values
attained for the orbits after their dynamic evolution while starting from zero after the total
time of simulation.  The $e_{\rm pl}$ variation is relatively large between 1~au and 2~au, roughly
coinciding with the stellar HZ.  In the region between 2~au and 4~au, the eccentricity variations remain
relatively low, thus indicating stable orbits.  The green color represents
the initial value $e_{\rm pl}=e_0$
and the other colors represent the $e_{\rm max}$ value the Earth-mass planets attained for the
respective choices of ICs in $e_{\rm pl}$ and $a_{\rm pl}$.  The colored circles
labeled e, b, c, and f depict the orbital locations and sizes of four of the five system planets
in the $a_{\rm pl}$ - $e_{\rm pl}$ phase space.  The observed resonances at 2.2~au, 2.8~au, 3.1~au, 3.8~au,
and 4.12~au are due to 55~Cnc-f and 55~Cnc-d.  Similar resonances have been reported by \cite{ray08}.
However, due to the increased resolution, they seem more prominent in this work.}
\label{fig:5}
\end{center}
\end{figure*}

\begin{figure*}
\begin{center}
  \includegraphics[scale=.4,angle=0]{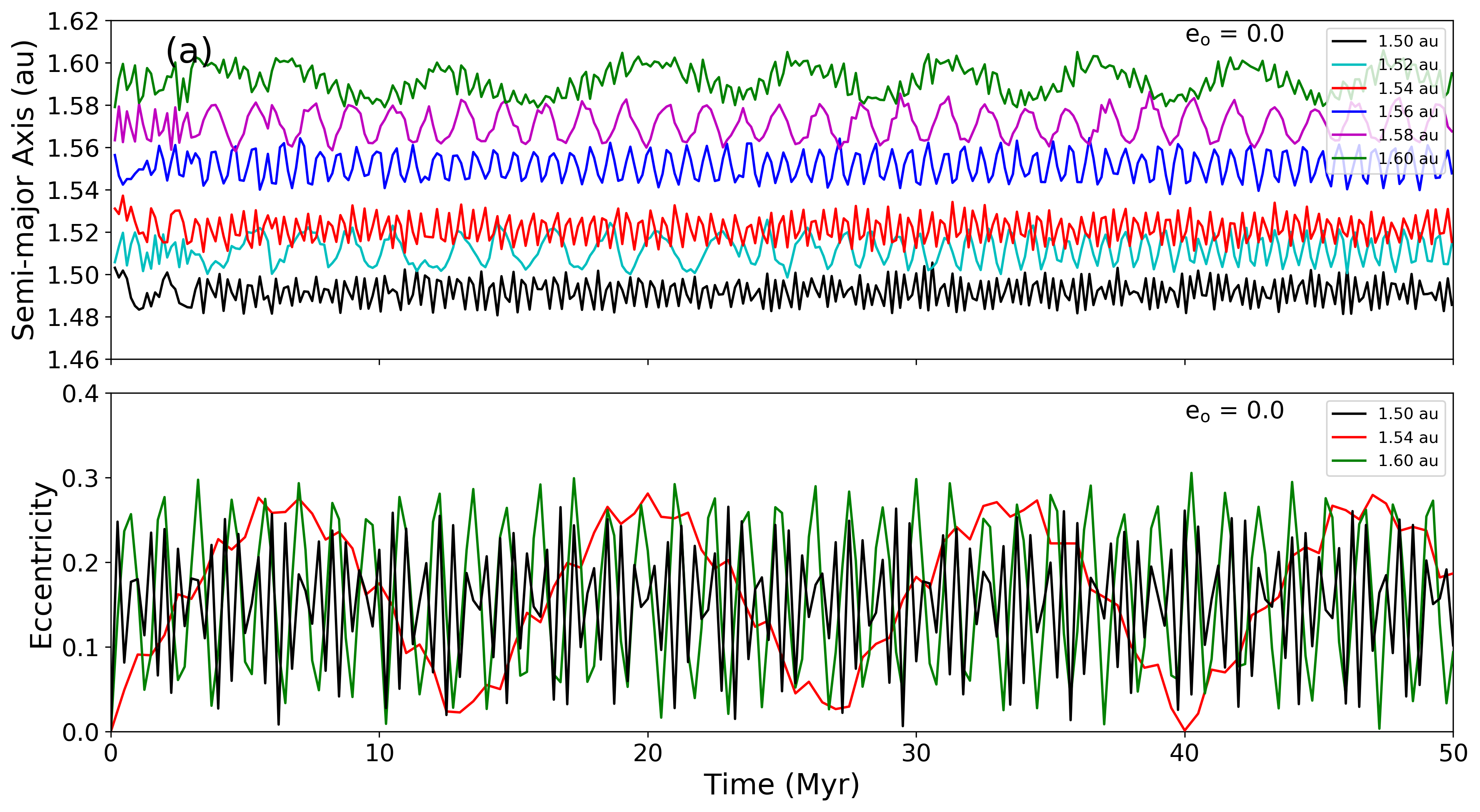} \\
  \includegraphics[scale=.4,angle=0]{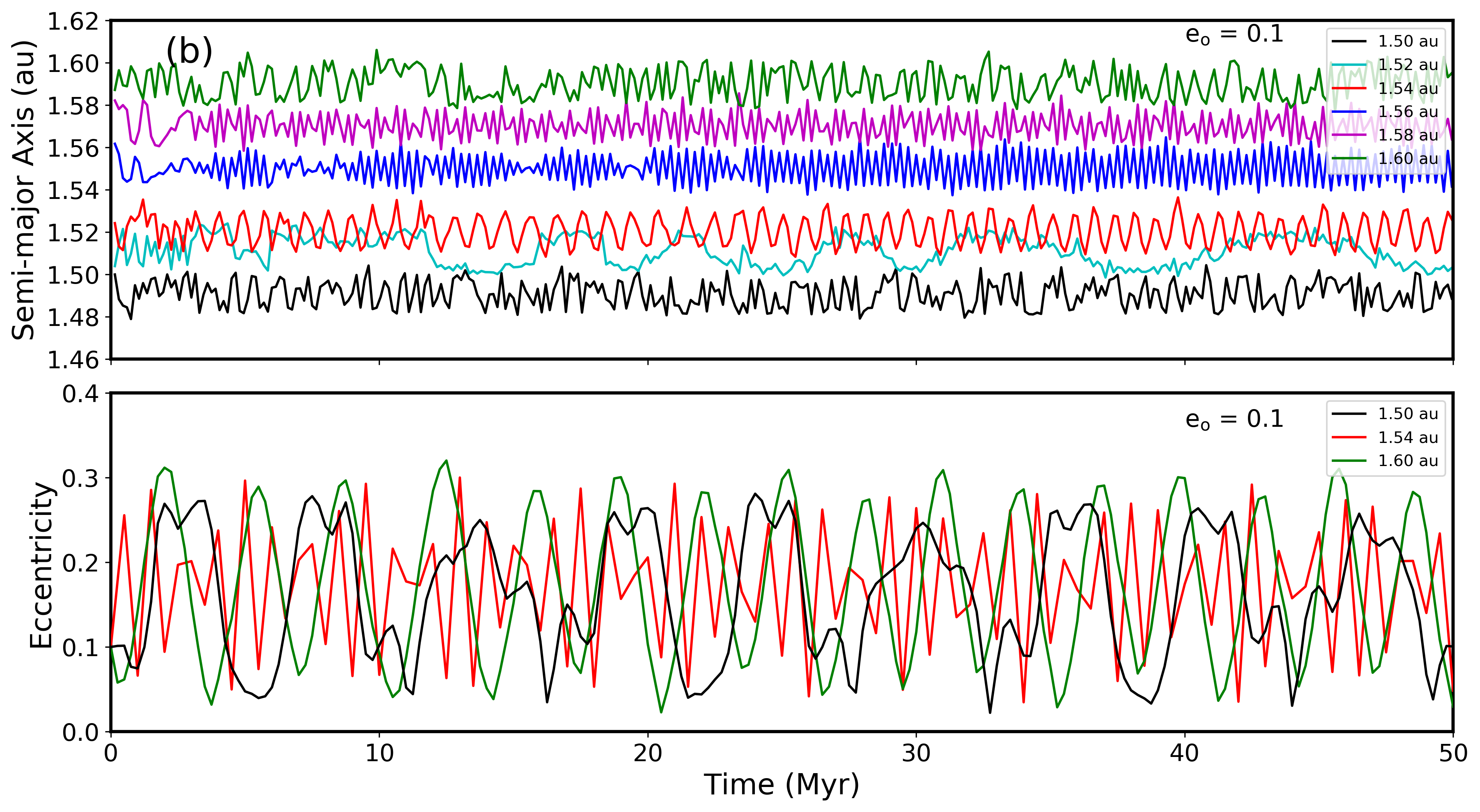} \\
  \includegraphics[scale=.4,angle=0]{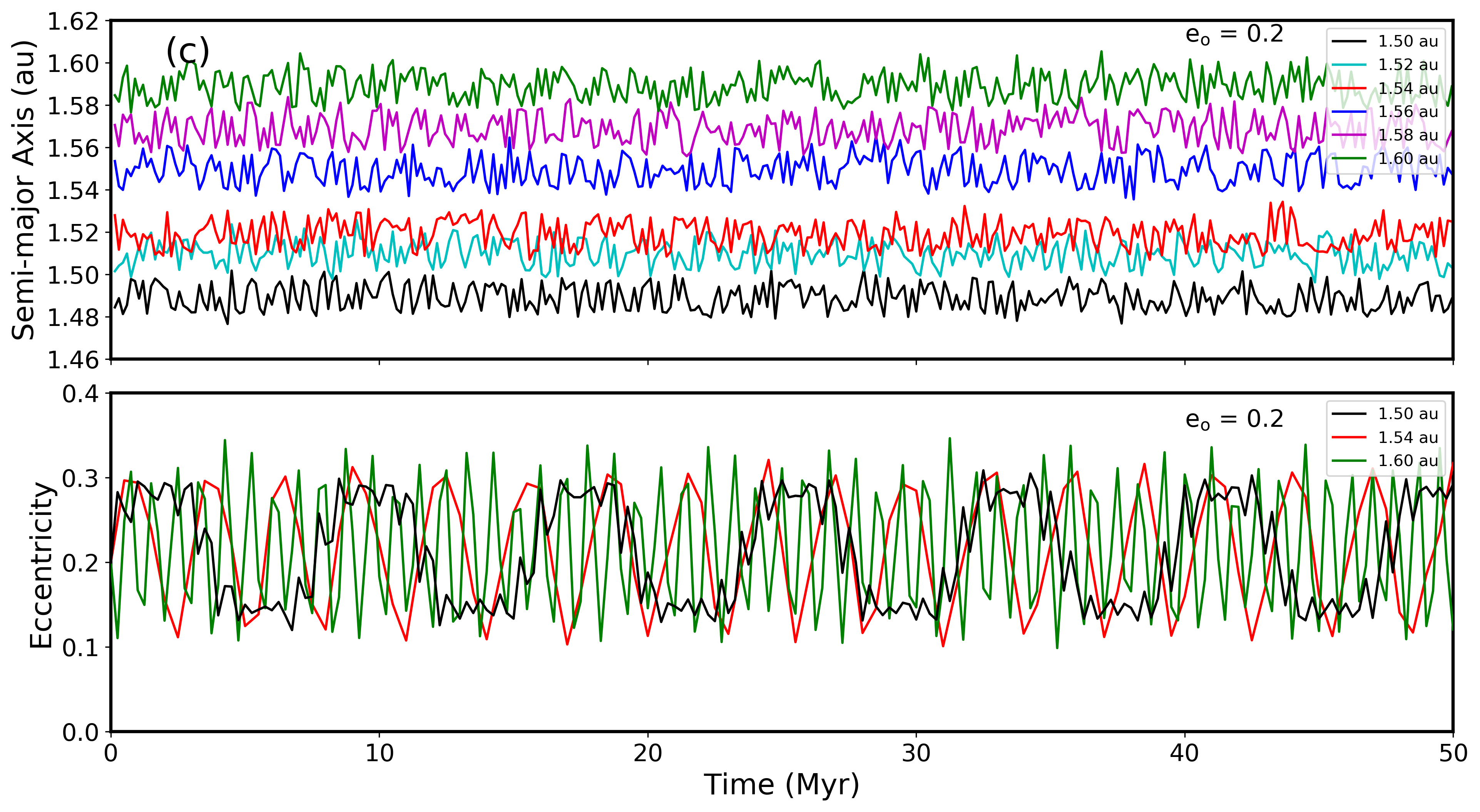}
\caption{Orbital simulations of Earth-mass test planets situated in the outskirts of the HZ (1.5~au and 1.6 au) based on the updated parameters from \cite{bou18}.
The simulation time (50 Myr) extends up to the onset of orbital stability. The time series plots for semi-major axes (top panel) and the eccentricities (bottom panel) indicate the time evolution of the test planets starting at $a_0$, as indicated, with the choices of $e_0$ given as 0.0, 0.1, and 0.2, respectively.  In each panel,
$a_{\rm pl}$ and $e_{\rm pl}$ are found to display stable orbits with only minor deviations from their initial values.}
\label{fig:6}
\end{center}
\end{figure*}

The choices of phase space and map resolution are similar to those considered in Section 3.1.  However, the ICs used to simulate the planetary orbits are now based on the best-fit values reported by \cite{bou18}, see Table 2.  The primary difference between their work and the previous work is given by the eccentricity of Cnc~55-f, which was updated from 0.32 to 0.08.  This lower value for the eccentricity results in a spatially increased dynamically stable region as given in Figs.~\ref{fig:5} and \ref{fig:6}, see discussion below.

Figures \ref{fig:5}a and \ref{fig:5}b show the stability region between 1~au and 2~au, which also largely coincides with the stellar HZ; see the $e_{\rm pl}$ - $a_{\rm pl}$ phase space.  The locations of the five system planets are shown as color-filled circles.  The $e_{\rm max}$ values (color coded) remain relatively low as shown in Fig.~\ref{fig:5}a for the added Earth-mass planets.  In the HZ, the $e_{\rm max}$ is less than 0.1 for relatively low initial eccentricities ($e_0$), thus indicating stable orbits.  The stability region extends up to $\sim$4~au for $e_0$ less than 0.2.  The unstable mean motion resonances (MMRs) due to the interactions with Cnc~55-f and Cnc~55-d reveal themselves in the phase space at around 1.65~au, 1.9~au, 2.0~au, 2.5~au, and 3.0~au. Despite these MMRs, the injected Earth-mass planets are expected to maintain stable orbits in the inner regions ranging from 1.0~au to 1.6~au.

In order to explore the time series evolution of the semi-major axis and the eccentricity of the added Earth-mass planet, we chose 33 different ICs in the stable phase space region as given in Fig. \ref{fig:5}b. Integrating fewer ICs is computationally less expensive, thus allowing integrations for longer orbital periods.
Therefore, the initial semi-major axis of the Earth-mass planet $a_0$ was varied from 1.50~au to 1.60~au with a step size of 0.01, and the initial eccentricity $e_0$ was varied from 0.0 to 0.5 with a step size of 0.1.  Each IC was simulated for 50 Myr.  These time series plots are shown in Fig.~\ref{fig:6}.  Figure \ref{fig:6}a depicts the evolution of $a_{\rm pl}$ and $e_{\rm pl}$ with $e_0$ set at 0.0.  To avoid overcrowding, only 6 and 3 points are plotted in the $a_{\rm pl}$ and $e_{\rm pl}$ panels, respectively.  The amplitude of the $a_{\rm pl}$ variation remains low for all the ICs for the total time of integration.  However, the $e_{\rm pl}$ variation is large, i.e., up to 0.3 for some values of $a_0$.  Furthermore, the orbits are periodic for up to 50 Myr.  Similarly, when $e_0$ was set to 0.1 and 0.2 (see Figs. \ref{fig:6}b and \ref{fig:6}c), the time series for $a_{\rm pl}$ and $e_{\rm pl}$ exhibit stable periodic orbits even though the amplitude of eccentricity oscillations varied by as much as 0.3.  For $e_0$ higher than 0.2, the system displayed unstable orbits within 50 Myr of simulation time.

The actual survival times of Earth-mass planets for various ICs and different system parameters, as given in Table~1 (earlier data) and Table~2 (data by \cite{bou18}), are given in Table 3.  These simulations pertain to circular orbits, i.e., $e_0=0.0$.  Here $a_0$ is set between 1.50~au and 1.60~au with a step size of 0.01~au.  For each combination of $a_0$ and $e_0$, the maximum eccentricity is recorded for the total time of simulation.  The results show that Earth-mass planets with $a_0$ between 1.50~au and 1.60~au based on the data of Table 1 did not survive (except for $a_0 = 1.60$~au).  However, when the updated data by \cite{bou18} are used, 
the Earth-mass planets survived for (at least) a 50~Myr timespan for all ICs.  The outcomes for the test planets, including their survival times, have directly been taken from the \texttt{mercury} output files.  This latter result is attributable to the updated value of the eccentricity for Cnc~55-f, now identified as close to circular.


\section{Comments on Previous Studies}

There had been previous studies on the possibility of Earth-mass planets around 55~Cnc, especially planets located in 55~Cnc's HZ, including work based on detailed orbital stability simulations.  For example, \cite{ray08} identified a large stable zone extending from 0.9 to 3.8~au at planetary eccentricities below 0.4.  This zone of stability encompasses a large fraction of 55~Cnc's OHZ, determined to extend from $\sim$0.59 to $\sim$1.43~au (see Sect.~3).  \cite{ray08} argued that the zone of stability may in fact contain 2 to 3 (albeit unobserved) additional planets.  In fact, based on the system data given by \cite{bou18}, we are largely able to confirm the previous results by \cite{ray08}.
This statement is based on a set of orbital integrations for Earth-mass test planets where we studied the time series evolution of their semi-major axis and eccentricity.

However, there is a bigger picture here.  As it turns out, the eccentricity of 55~Cnc-f is of pivotal importance
for the outcome of the simulations.  Previously, based on work by \cite{end12} and \cite{lig16}, that eccentricity
was determined as 0.32.   With this value of $e_{\rm pl}$ the planet's apocenter is at 1.03~au, thus located
within the previously identified domain for stability for possibly habitable terrestrial planets.  Hence, the
stability regime is pushed considerably further outward (i.e., as far as $\sim$2.2 au), which is beyond the
outer limit of the stellar HZ.  Thus, it is found that possible Earth-mass planets originally placed between
1.0~au and 4.0~au (a region also encompassing most of 55~Cnc's HZ) are unable to survive their total simulation
time of 10 Myr, a result impressively demonstrated in phase space and through eccentricity analysis.

The recent study by \cite{bou18} yields 0.08 as eccentricity for 55~Cnc-f, and possible Earth-mass planets
in the system's inner region are found to be orbitally stable for simulation times of 50~Myr.  Although it is beyond
the scope of this study to decide on the most realistic observational data for the planetary system of 55~Cnc,
the close-to-circular value for 55~Cnc-f's eccentricity appears to be most plausible.  Previously, it has been
shown that planetary eccentricities derived from radial velocity measurements can be seriously overestimated
\citep{she08,zak11}.  Additionally, both 55~Cnc-e (and 55~Cnc-f) are very unlikely to have high eccentricities
because the eccentricities of close-in planets are expected to be damped due to tidal dissipation \citep{bol13}.
The recent work by \cite{bou18} is highly comprehensive as it studies both the thermal and orbital properties
of the system planets in great detail; it also considers the impact of the stellar magnetic cycle on the results.

In a separate approach \cite{cun12} argued in favor of the existence of a possibly habitable planet at $\sim$1.5~au.  This opinion was based on an application of the Titius-Bode rule (TBR) to the system.  Despite some positive outcomes regarding the TBR, including work by \cite{bov13} who applied a generalized TBR to numerous exoplanetary systems (including those identified by {\it Kepler}), the applicability of the TBR continues to remain controversial\footnote{Previously, the nature of the TBR has been associated with planetary orbital stability, planet formation scenarios, or sheer numerology; see, e.g., \cite{lyn03} and \cite{cun12} for references and background information.  For example, \cite{hay98} concluded that
 ``[...] the significance of [the TBR] is simply that stable planetary systems tend to be regularly spaced [...]".}.  In the view of our study,
a planet-as-conjectured at $\sim$1.5~au is identified to be possible in consideration of the relatively small
orbital eccentricity of 55~Cnc-f.


\section{Summary and Conclusions}

The aim of this study is to explore the possible existence of Earth-mass planets
around 55~Cnc, an orange dwarf of spectral type G8~V, with an age somewhat larger
than the Sun.  Our main focus concerns the 55~Cnc's HZ, which assuming optimistic limits
extends between approximately 0.59~au to and 1.43~au.  55 Cnc is homestead of (at least) five planets
with masses ranging between super-Earth and Jupiter-type.  The planet closest to the stellar HZ
is 55~Cnc-f, with a semi-major axis of 0.77~au.  Furthermore, based on observational constraints,
there is a gap without planets between $\sim$0.8~au and $\sim$5.7~au.  Previously, it has been argued,
see, e.g., \cite{ray08}, that terrestrial planets may be able to exist within that gap,
including planets located in 55~Cnc's HZ.

Our study is based on detailed orbital stability simulations for hypothetical Earth-mass planets assuming
integration times of up to 50~Myr.  These planets are originally set between 0.0~au and 6.0~au.
We explore detailed phase-space maps and detailed time series analyses with focus on the evolution of
the eccentricity of the hypothetical Earth-mass planets.  As expected, those planets interact
gravitationally with the five system planets as well as the center star.  It is found that
the possibility for the existence of Earth-mass planets in the system, especially its HZ, strongly
depends on the adopted system parameters, notably the eccentricity of 55~Cnc-f.  Note that previously
both a high value ($e \sim 0.32$) and a low value ($e \sim 0.08$) have previously been deduced.
If the low value is adopted, the more plausible and most recent value, as given by \cite{bou18},
Earth-mass planets would be able to exist in the gap between 1.0~au and 2.0~au, thus implying the
possibility of habitable system planets.  We pursued test simulations for various parts of
the stellar HZ to verify this result.  If the high value for 55~Cnc-f's eccentricity is adopted,
typically, the Earth-mass test planets are found to collide with the star or the system planets,
or being ejected from the system.  This result is particularly evident for the region beyond 1.50~au,
which is most affected by 55~Cnc-f.

In conclusion, 55~Cnc should continue to be considered a favorable target for future habitable planet search missions.
Dynamical studies by \cite{sat17} for other systems have shown that Earth-mass planets can remain in stable
orbits for at least 1 Gyr, with a very small $a_{\rm pl}$ and $e_{\rm pl}$ deviations from their initial values.
Regarding the GJ~832 system, the planetary stability region was prominent and extended to a larger phase space,
i.e., up to 0.5 au, between the inner and outer planets \citep{wit14}.  Moreover, \cite{hin15} predicted
a possible third planet in the Kepler-47 circumbinary system by using the MEGNO based on long-term stability
simulations, which allowed identifying quasi-periodic planetary orbits.  In case of 55~Cnc, Earth-mass planets
in stable orbits are found to be possible in its HZ (and in various regions outside of that domain as well),
which is consistent with findings by \cite{ray08} and others.


\begin{ack}
This work has been supported by the Department of Physics, University
of Texas at Arlington.  We wish to thank the anonymous referee for their
comments which helped to improve the quality of our manuscript.
\end{ack}


\clearpage


\begin{table}
\tbl{Orbital Parameters of the 55 Cancri Planets, Previous Work
\label{tab:planets_param}}
{
\begin{tabular}{lccccccc}
\noalign{\smallskip}
\hline
\noalign{\smallskip}
Planet           & Mass         & $a$   & $e$ & $i$ & $\omega$ & $MA$ & Orbital Period \\
...              & ($M_\oplus$) & (au)  & ... & ($^\circ$) &($^\circ$) & ($^\circ$) & (days)   \\
\noalign{\smallskip}
\hline
\noalign{\smallskip}
55 Cancri e & 8.3700 & 0.0156 & 0.170 & 7.5 &  90 & 204.032 & 0.7365 \\
55 Cancri b & 264.75 & 0.1148 & 0.010 & 0.0 & 110 &  19.88  & 14.650 \\
55 Cancri c & 54.380 & 0.2403 & 0.005 & 0.0 & 356 & 116.10  & 44.364 \\
55 Cancri f & 57.209 & 0.7810 & 0.320 & 0.0 & 139 & 198.61  & 259.80 \\
55 Cancri d & 1169.6 & 5.7400 & 0.020 & 0.0 & 254 & 10.89   & 5169.0 \\
Earth-mass planet    & 1.0    & 0.0 $-$ 6.0 & 0.0 $-$ 0.4 & 0.0 $-$ 20 & 0.0 & 0.0 & ... \\
\noalign{\smallskip}
\hline
\noalign{\smallskip}
\end{tabular}
}
\begin{tabnote}
The planetary parameters (mass, $a$, $e$, $i$, $\omega$, and $MA$) of the 55 Cancri planets are listed,
see references \cite{fis08}, \cite{daw08}, \cite{win11}, \cite{end12}, and \cite{lig16}).  The
ascending node $\Omega$ is set to zero for all planets.
\end{tabnote}
\end{table}

\clearpage


\begin{table}
\tbl{Orbital Parameters of the 55 Cancri Planets, Data by \cite{bou18}.
\label{tab:planets_param_new}}
{
\begin{tabular}{lcccccc}
\noalign{\smallskip}
\hline
\noalign{\smallskip}
Planet           & Mass         & $a$   & $e$ & $i$        & $\omega$   \\
...              & ($M_\oplus$) & (au)  & ... & ($^\circ$) & ($^\circ$) \\
\noalign{\smallskip}
\hline
\noalign{\smallskip}
55 Cancri e & 7.99 & 0.0154 & 0.05 & 6.41 &  86 \\
55 Cancri b & 255.4 & 0.1134 & 0.00 & 0.0 & $-21.5$  \\
55 Cancri c & 51.2 & 0.2373 & 0.03 & 0.0 & 2.4  \\
55 Cancri f & 47.8 & 0.7708 & 0.08 & 0.0 & $-97.6$   \\
55 Cancri d & 991.6 & 5.957 & 0.13 & 0.0 & $-69.1$ \\
Earth-mass planet    & 1.0    & 0.0 $-$ 6.0 & 0.0 $-$ 0.4 & 0.0 $-$ 20 & 0.0  \\
\noalign{\smallskip}
\hline
\noalign{\smallskip}
\end{tabular}
}
\begin{tabnote}
The planetary parameters (mass, $a$, $e$, $i$, $\omega$) of the 55 Cancri planets are listed. The
ascending node $\Omega$ and the mean anomaly $MA$ are set to zero for all planets.
\end{tabnote}
\end{table}

\clearpage


\begin{table}
\tbl{Survival Times of Injected Earth-mass Planets for Circular Orbits \label{tab:survival_time}}
{
\begin{tabular}{lccccc}
\noalign{\smallskip}
\hline
\noalign{\smallskip}
$a_0$  & $e_0$   & $e_{\rm max}$ &  Survival Time         & Outcome 	 & Outcome				 \\
(au)               & ...                 & ...           &  (yr)                  & (Data of Table~1) & (Data of Table~2)    \\
\noalign{\smallskip}
\hline
\noalign{\smallskip}
1.50 & 0.0 & 0.62  &   107,554   & collided with f  &	survived 50 Myr	\\
1.51 & 0.0 & 0.95  &    93,264   & collided with d  &	survived 50 Myr	\\
1.52 & 0.0 & 0.63  &   752,416   & collided with f  &	survived 50 Myr	\\
1.53 & 0.0 & 0.76  &    69,005   & collided with f  &	survived 50 Myr	\\
1.54 & 0.0 & 0.65  &    23,998   & collided with c  &	survived 50 Myr	\\
1.55 & 0.0 & 1.20  &    18,115   & ejected          &	survived 50 Myr	\\
1.56 & 0.0 & 0.75  &    76,618   & collided with f  &	survived 50 Myr	\\
1.57 & 0.0 & 0.62  &    36,591   & collided with f  &	survived 50 Myr	\\
1.58 & 0.0 & 0.66  &   896,348   & collided with f  &	survived 50 Myr	\\
1.59 & 0.0 & 0.70  &   183,259   & collided with f  &	survived 50 Myr	\\
1.60 & 0.0 & 0.20  & 1,000,000   & survived (a$_{\rm final}$ = 1.64 au) &	survived 50 Myr	     \\
\noalign{\smallskip}
\hline
\noalign{\smallskip}
\end{tabular}
}
\begin{tabnote}
The indicated e$_{\rm max}$ and survival times (columns 3 and 4) refer to the data of Table~1
with the outcomes listed in column 5.  For the data set of Table~2, reflecting the work by \cite{bou18},
the survival status of the test planets is listed in column 6.  Results are obtained for specific
choices of the initial values $a_0$ and $e_0$.
\end{tabnote}
\end{table}

\clearpage

===========================================================

\clearpage


\clearpage


\clearpage


\emph{\clearpage}

\end{document}